\documentclass[iop]{emulateapj}
\usepackage{epsfig}

\def\mnras{Mon.\ Not.\ R. Astron.\ Soc.}
\def\apj{Astrophys.\ J.\ }
\def\aj{Astron.\ J.}

\def\aap{Astron.\&Astrophys}

\usepackage[usenames]{color}

\newcommand{\LCDM}{$\Lambda$CDM}

\shorttitle{The Cosmic Abundance of Classical Milky Way Satellites}
\shortauthors{Strigari and Wechsler}
\begin{document}

\title{The Cosmic Abundance of Classical Milky Way Satellites}
\author{Louis E. Strigari and Risa H. Wechsler}
\affil{Kavli Institute for Particle Astrophysics and Cosmology,
  Physics Department, Stanford University, Stanford, CA 94305 USA}

%TC:break Abstract
\begin{abstract}
  We study the abundance of satellites akin to the brightest,
  classical dwarf spheroidals around galaxies similar in magnitude and
  isolation to the Milky Way and M31 in the Sloan Digital Sky
  Survey. From a combination of photometric and spectroscopic
  redshifts, we bound the mean and the intrinsic scatter in the number
  of satellites down to ten magnitudes fainter than the Milky
  Way. Restricting to magnitudes brighter than Sagittarius, we show
  that the Milky Way is not a significant statistical outlier in its
  population of classical dwarf spheroidals. At fainter magnitudes, we
  find an upper limit of 13 on the mean number of satellites brighter
  than the Fornax dwarf spheroidal.  Methods to improve these limits that utilize full
  photometric redshift distributions hold promise, 
  but are currently limited by incompleteness at the very
  lowest redshifts.  Theoretical models are left to explain why the
  majority of dark matter subhalos that orbit Milky Way-like galaxies
  are inefficient at making galaxies at the luminosity scale of the
  brightest dwarf spheroidals, or why these subhalos predicted by
  \LCDM\ do not exist.
\end{abstract}
%TC:break _main_

\keywords{dark matter --- galaxies: dwarf --- galaxies: formation ---
  halo --- Local Group}

\maketitle

\section{Introduction} 
\label{sec:introduction}
Around the Milky Way (MW) Galaxy orbit nearly two dozen known
satellite galaxies that have a range of magnitudes down to twenty times
fainter than the MW itself. The two brightest satellites, the Large
Magellanic Cloud (LMC) and the Small Magellanic Cloud (SMC), are
approximately two and four magnitudes fainter than the MW,
respectively. Numerical simulations indicate that there is a $\lesssim
10\%$ chance that a $10^{12} M_\odot$, MW-mass dark matter halo
hosts two satellites as massive as the Magellanic
Clouds~\citep{BoylanKolchin:2009an,Busha:2010gi}. From an
observational perspective, analysis of galaxies from the Sloan Digital Sky Survey Seventh
Data Release (SDSS DR7) indicates that MW magnitude-galaxies have two
galaxies as bright as the Magellanic clouds only $\sim 5\%$ of the
time, and on average have $\sim 0.3$ satellites within four magnitudes
of the MW~\citep{Liu:2010tn,Guo:2011qc,Lares2011,Tollerud:2011wt}.
Thus the MW is somewhat atypical in having two satellites with this
luminosity difference, but the probability of having a given number at
this scale is in excellent agreement between theory and observation
\citep[see further discussion in][]{Busha:2010gi}.

Beyond the Magellanic Clouds (MCs), the brightest satellites are the
Sagittarius, Fornax, and Leo I dwarf spheroidals (dSphs),
approximately six, eight, and nine magnitudes fainter than the MW,
respectively. Though at present challenging, a measurement of the
abundance of satellite galaxies as bright as these classical dSphs is
important for several reasons.  From the observational perspective, it
improves our understanding of the MW and its place on a cosmic scale,
providing information on the number of bright satellites around other
galaxies over a regime in which we are believed to be complete around the
MW~\citep{Walsh2009}, in particular away from the Galactic plane 
~\citep{1997AJ....113..624K,2010AdAst2010E..21W}.
From the theoretical perspective, understanding
the abundance of classical dSphs is important because these objects
reside in the least massive dark matter halos that contain visible
light, corresponding to halo mass scales at which various processes
including supernova feedback effects and suppression of gas accretion
from reionization becomes
important~\citep{Bullock:2000wn,Benson:2001au,Somerville:2001km}.

Though the above processes certainly affect the formation of dSphs,
theoretical models that include them have still found it challenging
to match both the distribution of luminosities of the bright MW
satellites and their kinematic properties~\citep{Font:2011ds}.
Estimates of the luminosity function down to scales of the classical
dSphs predict that they reside in dark matter halos with velocity
dispersions $\gtrsim 20$ km/s~\citep{Cooper:2009kx}.  These models
typically predict tens of satellites brighter than Fornax, $\sim 10^7
L_\odot$. Both of these predictions are in tension with observations
of the MW satellite population. On the one hand, the predicted
velocity dispersion for bright satellites is nearly two times larger
than the observed $\sim 10$ km/s velocity dispersion of bright dSphs.
On the other hand, since the observational sample is complete for
objects with reasonable surface brightness with luminosity $\gtrsim
10^7 L_\odot$, it is difficult to invoke that a population of objects
this luminous has been missed by observations.

More detailed analysis of dSph kinematics indicates that their maximum
circular velocities are $\lesssim 30$ km/s ~\citep{Strigari:2010un}.
In particular, even though it is very bright, Fornax has a strongly
constrained maximum circular velocity at a relatively low $\sim 20$
km/s. Matching these results with numerical simulations of the
Galactic satellite population in a $\Lambda$CDM
cosmology~\citep{Springel2008,Diemand:2008in} indicates that on average there should
be $\sim 25-75$ of dark matter satellites with maximum circular
velocity greater than that of Fornax, that are either too faint to be
detected in surveys or devoid of baryonic material
entirely~\citep{BoylanKolchin:2011de}.

The above results indicate that from a combination of observations and
theory, the classical problem of the abundance of satellites within
\LCDM\ can be boiled down to a ``Fornax problem": more specifically,
why is it that for an observed galaxy at the luminosity scale of $\sim
10^7 L_\odot$, there are scores of dark subhalos that have the same
dark matter mass, but apparently no stars in them at all?  There are a
few outstanding ideas that remain to answer this question. First, the
tension may point directly to severe inefficiency and stochasticity of
galaxy formation at the dark matter halo mass scale of the dSphs.
Second, it may be that there are required modifications to the current
sample of numerical simulations: in particular, baryons may
significantly modify the dark matter distributions in satellites
(e.g. \citet{Wadepuhl2011,Parry:2011iz,DiCintio2011}), or there may even be a necessary
modification to the cosmological model~\citep{Lovell:2011rd}. The
third idea is perhaps the most straightforward of all; namely, that
the MW is rare amongst galaxies of its kind, a result of a rare
downward fluctuation in its population of bright satellites.

In this paper, we address for the first time this latter issue of the
distribution of satellites as bright as the classical dSphs around
MW-analog galaxies.  We construct an observational sample of
spectroscopic galaxies like the MW using SDSS, and search for faint
satellites using SDSS DR8 imaging data and photometric redshift
probability distributions.  At the faintest end, we place an upper
limit on the number of satellites down to the magnitude scale of the
Leo I dSph. At the brightest end, we find an average of $\lesssim 2$
objects brighter than Sagittarius. Our results imply that, in terms of
its bright dSph satellite population, the MW does not
stand out as a significant statistical outlier.

\section{Data} 
\label{sec:data}
Our analysis begins by selecting MW-analog galaxies, which is similar
to the analysis presented in~\citet{Liu:2010tn}.  We refer to these
galaxies as primaries.  We use a magnitude for the MW of
$M_V=-20.9$~\citep{vandenBergh2000}, which corresponds to
$M_{\rm r}=-20.4$~\citep[see][for discussion]{Liu:2010tn}, and we consider
primaries within $\pm 0.25$ magnitudes of this value.  We consider a
second set of slightly brighter galaxies, with $M_{\rm r}=-20.7$, for
comparison with M31.  Primary galaxies are selected from the NASA-Sloan
Atlas ({\tt v0\_1\_1})\footnote{http://www.nsatlas.org/data}, which
combines SDSS data for nearby galaxies ($z < 0.055$) with more complete
spectroscopy and imaging from GALEX.  We limit our analysis to
isolated galaxies that have no galaxy with a luminosity equal to or
greater than the MW or M31 within 0.4 Mpc (following the same methods
described in \citealt{Liu:2010tn}).  We make a few cuts to exclude
very nearby galaxies which may have incorrect photometry, specifically
those with $r < 11$, {\tt SIZE < 5*SERSIC\_TH50}, or with $g-r < 0$.

To perform the search for satellites around MW and M31-analogs, we
utilize the both DR8 imaging and spectroscopic
catalog~\citep{Aihara:2011sj}.  We use the galaxy catalog and the
photometric redshifts probability distributions
from~\cite{Sheldon:2011fm}, which is complete for $r < 21.8$. The
spectroscopic sample is complete for $r < 17.77$.  We exclude primary
galaxies for which the DR8 photometry is incomplete within 250 kpc of
the primary (using a mask kindly provided by Erin Sheldon; the
motivation for this cut will become more evident in
Sec.~\ref{sec:method}).

For a given magnitude difference from the primary, $\Delta m$, and
satellite absolute magnitude threshold, Table~\ref{tab:properties}
provides the corresponding apparent $r$-band magnitude threshold and
the number of primary galaxies. Columns 4-6 correspond to our
photometric sample, while columns 7-9 correspond to our spectroscopic
sample. The meaning of these samples will become more clear from our
discussion in Sec.~\ref{sec:method}.  For the photometric sample of
MW-analog primaries, we have explicitly included $\Delta m = 7.7$,
corresponding to the Fornax dSph.  In comparison to the sample
of~\citet{Liu:2010tn}, our sample with a threshold cut at the
magnitude of the brightest satellites, i.e. the MCs, is smaller
because we restrict to more nearby galaxies in the NASA-Sloan Atlas ($z <
0.055$).  Since the primary focus of our analysis is on the faintest
satellites, our conculsions are insensitive to the smaller sample of
very bright satellites.

%%%% Table of primary properties. 
%\begin{deluxetable*}{cccccccl}
%\tablecolumns{8}
%\tablewidth{0pc}
%\tablecaption{
%Properties of primary galaxies for each magnitude limit. The upper six rows correspond to 
%Milky Way-like primaries, while the bottom six rows correspond to M31-like primaries. 
%\label{tab:properties}}
%\tablehead{\colhead{Sample} & \colhead{$\Delta m$} & 
%\colhead{Sat. Abs. Mag.} &  \colhead{Primary App. Mag.} & \colhead{No. of Primaries} & \colhead{$\langle z \rangle$} 
%& \colhead{$z_{max}$} & $N(> \Delta m)$
%}
%\startdata
%MW &9&-11.9&12.8&148&0.011&0.016&$<100$\\
%&7.7&-13.2&14.1&1117&0.020&0.029&$<13$\\
%&7&-13.9&14.8&2727&0.026&0.039&$3.10 \pm 1.60$\\
%&6&-14.9&15.8&7574&0.038&0.055&$1.50 \pm 0.47$\\
%&5&-15.9&16.8&8388&0.040&0.055&$0.92 \pm 0.26$\\
%&4&-16.9&17.8&8388&0.040&0.055&$0.51 \pm 0.15$\\
%\hline
%M31&9&-12.2&12.8&176&0.013&0.018&$<60$\\
%&8&-13.2&13.8&958&0.020&0.030&$<13$\\
%&7&-14.2&14.8&3370&0.030&0.045&$2.41\pm1.55$\\
%&6&-15.2&15.8&7427&0.040&0.055&$1.31\pm0.49$\\
%&5&-16.2&16.8&7432&0.040&0.055&$1.02\pm0.47$\\
%&4&-17.2&17.8&7432&0.040&0.055&$0.50\pm0.21$
%\enddata
%\end{deluxetable*}
%%%%% End table of primary properties. 

%%% Table of primary properties. 
\begin{deluxetable*}{ccc||ccc|cccc}
\tablecolumns{8}
\tablewidth{0pc}
\tablecaption{
Properties of primary galaxies for each magnitude limit. The upper eight rows correspond to 
Milky Way-like primaries, the bottom seven rows correspond to M31-like primaries. 
\label{tab:properties}}
\tablehead{
\colhead{} &  \colhead{} & \colhead{} & \multicolumn{3}{c}{Photometric} &  \colhead{} & \multicolumn{3}{c}{Spectroscopic} \\
\cline{4-6} \cline{8-10} \\
\colhead{Sample} & \colhead{$\Delta m$} & 
\colhead{Satellite Absolute} &  \colhead{Primary Apparent} & \colhead{Number of} & \colhead{Mean Number} & 
\colhead{}  & 
 \colhead{Primary Apparent} & \colhead{Number of} &  \colhead{Mean Number} \\
 \colhead{} & \colhead{} & \colhead{Magnitude} &  \colhead{Magnitude}& \colhead{Primaries} 
 & \colhead{of Satellites} & \colhead{} & \colhead{Magnitude} &\colhead{Primaries} & \colhead{of Satellites}
}
\startdata
MW &10&-10.9&11.8&49&$<400$&&--&--&--\\
&9&-11.9&12.8&148&$<100$&&--&--&--\\
&8&-12.9&13.8&727&$<30$&&--&--&--\\
&7.7&-13.2&14.1&1117&$<13$&&--&--&--\\
&7&-13.9&14.8&2727&$3.10 \pm 1.60$&&--&--&--\\
&6&-14.9&15.8&7574&$1.50 \pm 0.47$&&11.8&49&$1.42\pm0.56$\\
&5&-15.9&16.8&8388&$0.92 \pm 0.26$&&12.8&148&$0.62\pm0.18$\\
&4&-16.9&17.8&8388&$0.51 \pm 0.15$&&13.8&727&{\bf $0.25\pm0.05$}\\
\hline
M31&10&-11.2&11.8&41&$<300$&&--&--&--\\
&9&-12.2&12.8&151&$<60$&&--&--&--\\
&8&-13.2&13.8&834&$<13$&&--&--&--\\
&7&-14.2&14.8&3370&$2.41\pm1.55$&&--&--&--\\
&6&-15.2&15.8&7427&$1.31\pm0.49$&&11.8&41&$1.01\pm0.54$\\
&5&-16.2&16.8&7432&$1.02\pm0.47$&&12.8&151&$0.37\pm0.12$\\
&4&-17.2&17.8&7432&$0.50\pm0.21$&&13.8&834&{\bf $0.20\pm0.04$}
\enddata
\end{deluxetable*}
%%%% End table of primary properties. 

\section{Methods}
\label{sec:method}

Consider a galaxy that has been selected as a MW or M31-analog via aforementioned 
methods. We define the ``signal" region as circular area centered
around the primary, corresponding to a physical radius $R$ at the
redshift of the primary. For the main analysis in this paper we take a
$R = 250$ kpc for the signal region, corresponding to the approximate 
viral radius of the MW and M31~\citep{Springel2008,Diemand:2008in}. 
In the signal region we search for galaxies that are between two magnitudes fainter
and a threshold of $\Delta m$ magnitudes fainter than the primary. For
the $\imath^{th}$ primary, we label the number of galaxies
in the signal region as $n_{t,\imath}$. 
For comparison to the number of counts within the signal region, we 
associate with each primary a background region. The background region 
is  an annulus with an inner radius $R$, and an outer radius chosen to enclose 
the same area as the signal region. We label the number of galaxies
within the background region associated with the $\imath^{th}$ primary as
$n_{b,\imath}$. A local estimation of the background
via an annulus connected to the signal region has been shown to be unbiased,
even accounting for galaxy clustering~\citep{Chen:2005sg,Liu:2010tn,Lares2011,Guo:2011qc}. 

We are interested in obtaining an estimate for the mean number of satellites 
around MW and M31-like galaxies, given the measurements of $n_{t,\imath}$
and $n_{b,\imath}$ around a large sample of primaries. 
We obtain three separate estimates for the combination of $n_{t,\imath}$
and $n_{b,\imath}$,
which differ both in the cuts that are made on the data sample and the
method in which the sample is obtained. 
The first estimate uses spectroscopic redshift information for the satellites, 
while the second and third methods use photometric redshifts. 
We now detail specifically how each of these estimates are obtained in turn.   

\subsection{Method 1: Spectroscopic satellites} 
Our first method for estimating $n_{t,\imath}$ and $n_{b,\imath}$ uses
spectroscopic redshifts for both primaries and satellites.  Since the
DR8 sample is spectroscopically-complete down to $r<17.77$, we are
able to estimate the number of satellites brighter than $\Delta m =
6,5,4$.  As indicated in Table~\ref{tab:properties}, these magnitude
differences correspond to primary apparent magnitudes $r < 11.8, 12.8,
13.8$, respectively.  Specifically we find 49, 148, and 727 primaries
that satisfy $r < 11.8, 12.8, 13.8$, respectively.  For a given set of
primaries, we select galaxies in the spectroscopic sample that satisfy
two criteria. First, as described above we determine those galaxies in
the signal region that satisfy $R < 250$ kpc, as well as galaxies
within the corresponding background annulus.  Second, we impose a cut
so that both the galaxies in the signal region and the galaxies in the
background region lie within a redshift $\Delta z = 0.001$ ($300 {\rm
  km s}^{-1}$) of the primary. This redshift cut is appropriate when
accounting for both the expected physical size of the dark matter
halos of the primaries, and for redshift space distortions.

Using spectroscopic redshifts to identify satellites clearly reduces
background contamination from galaxies at vastly different redshifts
than the primary. However, the obvious downside is that, when
demanding the satellites have a spectroscopic redshift, the sample of
primaries is much smaller than the corresponding sample of primaries
obtained by only demanding complete photometry for the satellites.
Further, when restricting to spectroscopic satellites we run out of
very bright primaries, meaning we are unable to probe very faint
magnitude differences, $\Delta m > 6$, that are the main goal of this
work.

\subsection{Method 2:  Full photometric redshift distributions} 
For our second estimate of $n_{t,\imath}$ and $n_{b,\imath}$, we 
use the photometric galaxy sample of~\citet{Sheldon:2011fm}, which is 
complete down to $r < 21.8$. In this case, 
we are able to estimate the number of satellites 
down to magnitudes as faint as $\Delta m = 10$, corresponding to primaries with $r < 11.8$. The 
total number of primaries for each $\Delta m$ is shown under the photometric heading 
in Table~\ref{tab:properties}. 

For each galaxy with $r < 21.8$,~\citet{Sheldon:2011fm} assign a
probability, $p(z)$, that it resides in one of $N_{phot} = 35$
redshift bins. Method 2 specifically uses the information in these
$p(z)$ distributions. With these distributions in hand, again we start
by considering a single primary galaxy. We use the same redshift bins
in which the $p(z)$'s are calculated, and from these locate the
redshift bin that contains the primary. As described above we again
search for galaxies within the signal region $R < 250$ kpc. Given this
set of galaxies in the signal region, we then sample each of their
corresponding discrete $p(z)$ distributions.  This provides us with a
redshift for each galaxy within the signal region.  From a single
sampling of each $p(z)$ distribution, we determine the total number of
galaxies that fall into the redshift bin that contains the primary.
To obtain a statistically robust estimate of the total signal and
background probability distributions, we repeat the sampling procedure
$\sim 100$ times to obtain the mean number of counts in the
signal region in the same redshift bin as the primary, $\langle
n_{t,\imath} \rangle$.  Then following a similar reasoning as above,
we repeat this procedure for the annular background region to obtain
the corresponding mean number of counts in the redshift bin associated
with the background for each primary, $\langle n_{b,\imath} \rangle$.

As long as the photometric redshift distribution for each galaxy is
unbiased, the estimate of the number of satellites obtained from this
method is expected to be just as accurate as the estimate that uses
spectroscopic redshifts.  However, systematics in photometric redshift
information may bias the results. Of particular relevance is the fact
that the training set used in~\citet{Sheldon:2011fm} to obtain the
photometric redshifts may not have a fully representative sample of
faint, low redshift galaxies. This is precisely the sample of galaxies
that we are interested in identifying as satellites.  If the training
sample is underrepresented at low redshift, this will likely lead to a biased
$p(z)$ estimate for potential satellites, and underestimate of the
satellite number associated with low redshift hosts.  As shown below,
a comparison of the results obtained from this method to those
obtained from the spectroscopic sample (method 1), over the regimes in
which both are complete, may provide an estimate of the magnitude
regime in which photometric redshifts are not representative for
faint, low redshifts galaxies.

%%%% Full pz figure
\begin{figure*}
\begin{center}
\begin{tabular}{c}
 {\includegraphics[height=6.0cm]{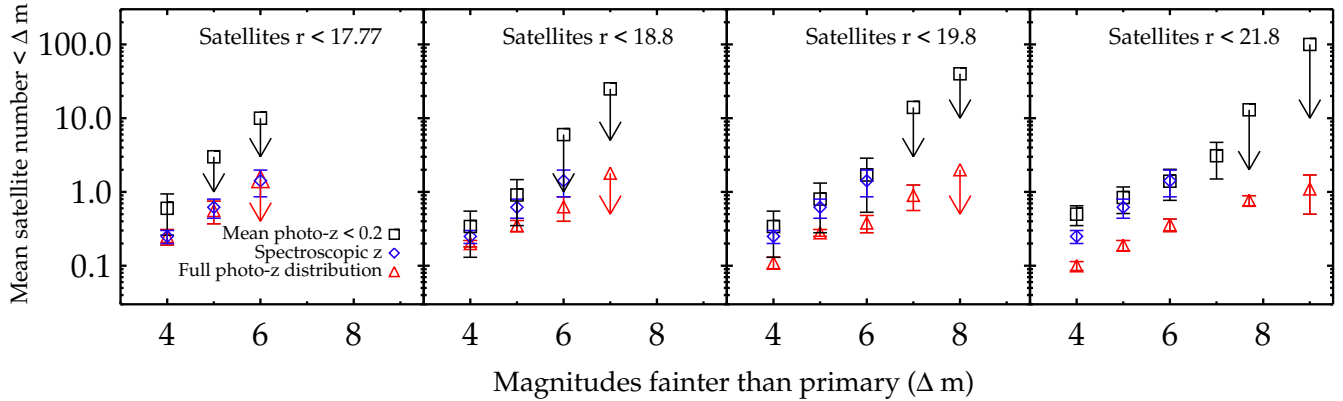}} \\
% {\includegraphics[height=4.2cm]{comparison_17.77.eps}} & 
 % {\includegraphics[height=4.2cm]{comparison_18.8.eps}} 
 % {\includegraphics[height=4.2cm]{comparison_19.8.eps}} & 
% {\includegraphics[height=4.2cm]{comparison_21.8.eps}} \\
\end{tabular}
\end{center}
\caption{Mean number of satellites within $\Delta m$ of MW-like primaries, 
using the three different methods described in Sec.~\ref{sec:method}. In all panels, 
blue diamonds include only satellites with spectroscopic redshift within $\Delta z = 0.001$ of the 
primary (method 1), and red triangles sample the full photometric redshift probability distributions
(method 2). In the three left panels, black squares only 
include satellites with mean photometric redshift less than 0.2, while in the right-most panel 
black squares only include satellites with mean photometric redshift less than 0.5 (method 3).
In each of the four panels, the magnitude limit on the population of satellites from 
the photometric analyses (methods 2 and 3) are indicated. 
In all panels, upper limits are indicated as downward arrows. 
\label{fig:fullpz}
}
\end{figure*}
%%%%

\subsection{Method 3: Mean photometric redshifts} 
Our third method to estimate $n_{t,\imath}$ and $n_{b,\imath}$ also uses information contained in the photometric redshift distributions. 
However, rather than directly sampling the full $p(z)$ distributions as above, this method simply 
uses the information contained in the mean of these distributions. 
Once a mean photometric redshift is assigned to a galaxy, 
implementation of this analysis method boils down to determining an appropriate redshift cut 
on galaxies included in the signal and background region.

To motivate the appropriate redshift cut, define the mean photometric redshift for a galaxy as 
\begin{equation} 
\langle z_p \rangle \equiv \sum_{\imath=1}^{N_{phot}} z_\imath p(z_\imath), 
\label{eq:meanz}
\end{equation} 
where $p(z_\imath)$ is the probability for the $\imath^{th}$ redshift bin. 
Though it is simple, ascribing to a galaxy a single redshift based upon Eq.~\ref{eq:meanz}
will typically overestimate the true redshift of the galaxy.
More specifically we are interested in the mean photometric 
redshifts for satellites that span the redshift range of our primary galaxies; 
this corresponding redshift range is $0.055$.
Examining the faintest sample of training set galaxies, $20.8 < r < 21.8$, over this redshift range of the 
primaries, we find that the ratio of the mean photometric redshift to the true redshift, $z_{true}$,  spans 
a wide range $\langle z_p \rangle/z_{true} \sim 3-30$. Further, even for the brightest sample of training 
set galaxies fainter than the spectroscopic completeness limits, $17.8 < r < 18.8$, we find 
$\langle z_p \rangle/z_{true} \sim 1-10$. 

In spite of the fact that typically $\langle z_p \rangle/z_{true} \gtrsim 1$, 
it is still possible to use the information on the mean photometric redshifts, provided that 
we place an appropriate upper cut on the mean photometric redshift of a galaxy that is 
allowed in our sample. In other words, we must
estimate the number of true satellites that are lost from the sample 
when including galaxies less than a given $\langle z_p \rangle$. 

We obtain this estimate for the loss fraction by again considering training set galaxies over the redshift range
spanned by the set of primary galaxies of interest. Specifically, out to the maximum primary redshift, 
we examine the distribution of $\langle z_p \rangle$ for training set galaxies that are brighter than 
a given apparent magnitude. 
Then from this distribution we determine the value of $\langle z_p \rangle_{90}$, which
we define as the redshift below which 90\% of the galaxies reside. We determine
$\langle z_p \rangle_{90}$ for each apparent magnitude threshold cut. For example we find that when 
searching for satellites down to $r < 21.8$, $\langle z_p \rangle_{90} = 0.5$. Further, 
we find that for $r < 19.8$, $\langle z_p \rangle_{90} = 0.2$. These are the cuts that
we utilize in our analysis below. 
Because of the large sample of primaries available when using
the photometric catalogue, we find that the results of method 3 are generally insensitive
to the chosen value for $\langle z_p \rangle_{90}$.

\subsection{Likelihood} 

With the methods outlined for determining the data sample of galaxies in the signal and background regions,
it remains to construct the likelihood for the data.
For a total sample of $n_p$ primaries, the mean number of satellites is the difference between the mean of the signal 
and background counts, 
\begin{equation}
\mu_s = \frac{1}{n_p}\sum_{\imath=1}^{n_p} n_{t,\imath} 
-  \frac{1}{n_p}\sum_{\imath=1}^{n_p} n_{b,\imath} = \mu_t - \mu_b. 
\label{eq:mean} 
\end{equation}
We define the variance of the signal and background distributions as $\sigma_t^2$ and $\sigma_b^2$. 
Assuming that the number of galaxies around a primary and the background are uncorrelated, the 
variance is 
%\begin{equation}
$\sigma_s^2 = \sigma_t^2 - \sigma_b^2$. 
%\label{eq:variance} 
%\end{equation}
Defining the parameter set that we estimate from the data as $\vec x  = [\mu_s,\sigma_s,\mu_b,\sigma_b]$, 
for a given magnitude threshold $\Delta m$, the probability for the mean and variance is 
\begin{eqnarray}
P(\vec x | \Delta m ) \propto \prod_{\imath=1}^{n_p}  
&\frac{1}{\sqrt{ \sigma_s^2 + \sigma_b^2}}&
\exp \left[-\frac{(n_{t,\imath}  - \mu_s-\mu_b )^2}{2(\sigma_s^2 + \sigma_b^2)} \right] \nonumber \\
\times &\frac{1}{\sqrt{\sigma_b^2}}&
\exp \left[-\frac{(n_{b,\imath} - \mu_b )^2}{2 \sigma_b^2} \right]. 
\label{eq:like} 
\end{eqnarray} 
The probability distribution $P(\vec x | \Delta m )$ can be   
thought of as the probability for the number of counts in the signal region, the first term in Eq.~\ref{eq:like}, 
weighted by a prior given by the number of background counts, the second term in Eq.~\ref{eq:like}. 

Equation~\ref{eq:like} is a general formula that can be used to estimate the mean and the variance of the 
satellite probability distribution, independent of the number of primaries that contribute to the sample. 
We integrate Eq.~\ref{eq:like} and marginalize over the background mean and variance to obtain the probability for 
$\mu_s$ and $\sigma_s$. In our results below, error bars are derived as the area
centered on the mean containing 68\% of the cumulative probability distribution for $\mu_s$. 
We use uniform priors on the parameters that we estimate, $P(\vec x) =$ const. 

Assuming that the respective number counts $n_{t,\imath}$ and $n_{b,\imath}$ for a given primary reflect 
the mean of an underlying redshift probability distribution, Eq.~\ref{eq:like} follows directly 
from the central limit theorem. This is explicitly true in the case of method 2, which 
makes the replacements $n_{t,\imath} \rightarrow \langle n_{t,\imath} \rangle$ and 
$n_{b,\imath} \rightarrow \langle n_{b,\imath} \rangle$ in Eq.~\ref{eq:like}. 
In the limit of a large number of primaries, we have verified that Eq.~\ref{eq:like}
reproduces the true mean of the full satellite probability distribution, as determined in 
\citet{Liu:2010tn}. 

%%%%Figure template 
\begin{figure*}
\begin{center}
\begin{tabular}{cc}
{\includegraphics[height=8.cm]{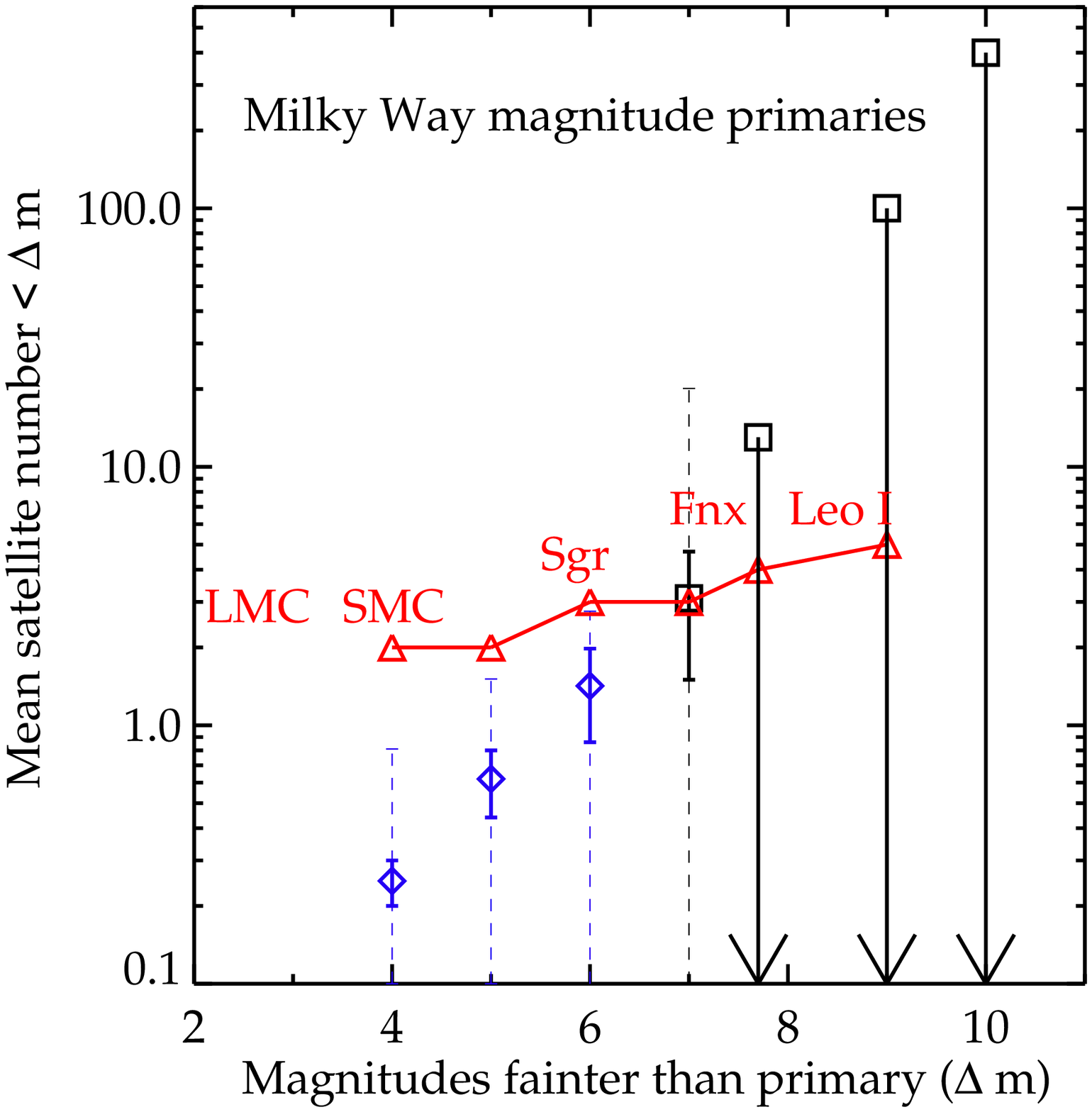}} & 
 {\includegraphics[height=8.cm]{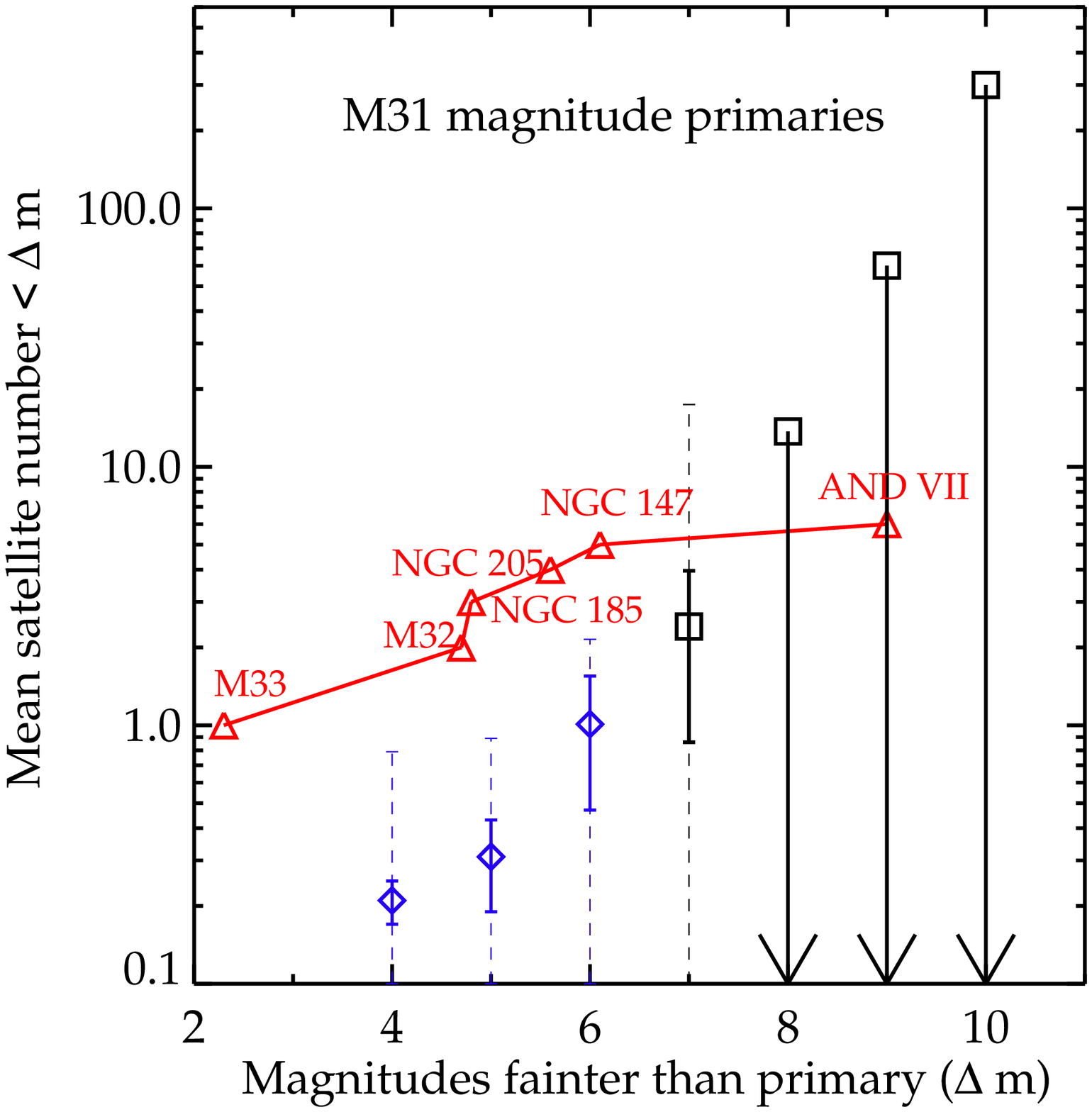}} \\
\end{tabular}
\end{center}
\caption{
{\it Left}: Mean number of satellites brighter than $\Delta m$ magnitudes fainter
than the primary galaxy, assuming primaries within $\pm 0.25$ magnitudes of the Milky Way. 
Blue diamonds are determined from the spectroscopic sample of satellites (method 1), black squares
from the photometric sample (method 3). The solid errors are the uncertainty on the mean, the
thin, dashed errors are the intrinsic scatter ($\sigma_s$ from Eq.~\ref{eq:like}). 
The arrows indicate 90\% c.l. upper limits. The red triangles indicate the
Milky Way satellites. {\em Right}: Same as left, except for primaries within $\pm 0.25$ magnitudes of M31. 
\label{fig:mean}
}
\end{figure*}
%%%%

\section{Results}
\label{sec:results}
We begin by comparing the results of the various methods for
estimating the number of satellites for several samples of host
galaxies.  The four panels in Figure~\ref{fig:fullpz} show the mean
number of satellites around MW-like primaries that are brighter than
$\Delta m$ magnitudes fainter than the primary, $\mu_s(< \Delta m)$, for
the three methods described above.  In each of the four panels, there
is a different threshold cut on the magnitude of signal and background
galaxies for the two methods that use photometric redshifts (methods 2
and 3).  From left to right, the magnitude cuts on signal and
background galaxies are $r< 17.8, 18.8, 19.8, 21.8$.  Results from the
spectroscopic analysis (method 1) are shown as blue diamonds in all
panels; by definition this method only includes galaxies in the signal
and background regions with $r < 17.77$.  Results from the full $p(z)$
sampling analysis (method 2) are shown as red triangles, and from the
cut on mean photometric redshifts (method 3) are shown as black
squares.

Figure~\ref{fig:fullpz} indicates that for satellites with $r < 17.8$
(leftmost panel), we find good agreement between all three methods,
with large error bars for the method using simple $p(z)$ cuts.  For
these primaries, the $p(z)$ distribution is determined by galaxies in the
main sample of SDSS, so the training sample is expected to be fully
representative.  The agreement between methods 1 and 2 is maintained
for $\Delta m = 4$ and $r < 18.8$ (left two panels).  However, we find
that this agreement weakens for $\Delta m = 5,6$, where the mean
abundances as determined from method 2 underestimate the results from
the spectroscopic analysis. This underestimation is most evident in
the right panel of Fig.~\ref{fig:fullpz}, and as discussed above most
likely indicates that galaxies with $r > 18.8$ and $z \lesssim 0.055$
are not adequately represented in the training sets.  In addition, as
expected, method 3 tends to overestimate the mean abundance for all
$\Delta m$ and all magnitude limits. This is true both in the regime
in which $\langle z_p \rangle_{90} = 0.2$, for satellites with
$r<19.8$ (left three panels), and $\langle z_p \rangle_{90} = 0.5$,
and for satellites with $r<21.8$ (right panel).  These results
indicate that the photometric redshift distributions for satellites
dimmer than $r=17.8$ are likely biased towards higher redshifts, as a
result of increasing incompleteness in the very lowest redshift
galaxies in the training sample for dimmer galaxies.

What leads to this bias in the photometric redshifts, and is it
avoidable?  First, it is important to realize that our regime of
interest is {\em very} low redshift, $z < 0.02$, well outside the
redshift range where photometric redshifts are generally used and
tested.  The $p(z)$ distributions calculated by \cite{Sheldon:2011fm}
use training sets from SDSS~\citep{Aihara:2011sj}, PRIMUS~\citep{Coil:2010df},
zCOSMOS~\citep{Lilly:2006va}, 2SLAQ~\citep{Cannon2006}, 
VVDS~\citep{Garilli2008}, DEEP2~\citep{Weiner:2004qm},
CNOC2~\citep{Yee:2000qj}, CFRS~\citep{Lilly1995}, 
and TKRS~\citep{Wirth:2004tm}. 
For galaxies in our main regime of interest,
$17.8 < r < 21.8$, the training sample is dominated by PRIMUS
\citep[][, Cool et al, in preparation]{Coil:2010df}, which covers more
than 9 sq. degrees to a depth of $i_{\rm AB} \sim 23.5$ (5.2
sq. degrees are included in the current training set).  
%In general,
%the spectrographs used for these redshift surveys are designed for
%higher redshifts, and they do not for example have a wavelength window
%that covers the 4000~\AA break in our regime of interest, $z < 0.02$,
%and thus can be unreliable in this regime.  
For PRIMUS, which is focused on science at $z > 0.2$, there are at
least two distinct issues that impact our analysis.  First, the
4000~\AA break falls out of the wavelength regime for $z < 0.2$;
without this feature the low-redshift spectra were more likely to be
assigned a lower confidence flag ($Q=3$) and were excluded in the
$p(z)$ estimation (Carlos Cunha, private communication).  Second, the
analysis pipeline does not even attempt to measure accurate redshifts
in the regime of our primaries ($z < 0.02$) because of the limited
velocity resolution of the low-dispersion prism used by PRIMUS (John
Moustakas \& Alison Coil, private communication).  Both of these will
impact the completeness of the low redshift sample and are likely to
bias the $p(z)$ distribution against the redshift range of our primary 
samples.

One might also consider whether the biased $p(z)$ distribution could
be avoided with another photometric redshift code.  However, this
range is challenging for any algorithm.  For example, template-fitting
codes without priors have known failure modes at very low redshift
\citep{Mandelbaum:2007dp}, and this would likely contaminate the low
redshift sample.

From the information in Figure~\ref{fig:fullpz} we are able to obtain
our best estimate for $\mu_s(<\Delta m)$ over the entire range of $\Delta
m$. Specifically for our best estimate of $\mu_s(<\Delta m)$ we use the
spectroscopic results for $\Delta m \le 6$, and the results from
method 3 for $\Delta m > 6$. Figure~\ref{fig:mean} shows these as the
main results of our analysis, in comparison to the observational points
from the MW and M31. The results from method 3 are explicitly shown in
the 6th column of Table~\ref{tab:properties}, and the results from
method 1 are shown in the last column of Table~\ref{tab:properties}.
Note that for the cases with a small number of primaries, which for
our MW sample corresponds to $\Delta m > 7$, only an upper limit on
the mean can be determined.

~\citet{Lares2011} report that the satellite population is uncertain
in the central 100 kpc around primary galaxies due to contamination
from features in the extended halos of primaries. We have directly
examined the images of many nearby, bright primaries and find that
generally this systematic is accounted for in
the~\citet{Sheldon:2011fm} galaxy sample used in our analysis. In
addition we have explicitly examined how our results are affected when
excluding the 100 kpc around primaries, and we generally find minimal
changes relative to those results presented in Fig.~\ref{fig:mean}.
For example, for Fornax-magnitude satellites around MW-like primaries,
we find that the 90\% c.l. upper limit on the mean reduces from 13 to
9 when excluding the central 100 kpc.

In addition to the distribution of the mean number of satellites, from
our likelihood analysis we are able to estimate the intrinsic scatter,
or $\sigma_s$, for each magnitude bin.  As for the mean, $\sigma_s$ is
determined from the full probability density distribution using
Eq.~\ref{eq:like}. As an example for MW-like primaries and $\Delta m =
[4, 5]$, we find a mean intrinsic scatter of $\sigma_s = [0.56 \pm
0.04,0.89 \pm 0.19]$, where the errors represent one-sigma
uncertainties as above. The best-fitting values for $\sigma_s$ are
shown as thin, dashed error bars in Fig.~\ref{fig:mean} for $\Delta m
\le 7$. Via the method outlined in~\citet{Liu:2010tn}, we are also
able to estimate the full probability distribution down to $\Delta m =
5$; here we find that the probability to obtain $[0,1,2,3,4]$ satellites
with $\Delta m < 5$ is $[0.59,0.25,0.11,0.03,0.02]$. Down to fainter
magnitudes, the spectroscopic sample is too sparse to measure the full
satellite probability distribution.  These results indicate that there
is still substantial intrinsic scatter in the satellite population,
even at the brightest scales.

We note that the limits we present are strictly valid over the regime
of surface brightness where the SDSS DR8 data is complete. For
galaxies with half-light surface brightness $\lesssim 22.5$
mag/acrmin$^2$, which is the surface brightness of Leo I, the SDSS
spectroscopic data is $\sim 90\%$ complete ~\citep{Blanton2005}. For
galaxies with surface brightness similar to Fornax or Sculptor,
$\lesssim 23.5$ mag/acrmin$^2$, the surface brightness completeness is
$\sim 50\%$.  We note that the surface brightness incompleteness has
not been estimated directly for the photometric sample, but the
agreement our results between the two sets of samples indicates that
the issues may be of similar magnitude.  We can get an estimate as to
how the surface brightness incompleteness affects our results by
comparing the integrated luminosity function of~\citet{Blanton2005}
that is corrected for incompleteness, as compared to the measured
luminosity function. Down to the magnitude of Fornax, for example we
find that the luminosity functions differ by a factor $\lesssim 2$
going to down surface brightness of 24 mag/acrmin$^2$. If interested
in constraining the population of objects down to this surface
brightness, this factor should be taken as a conservative correction
to the limits that are presented in Fig.~\ref{fig:mean}.  Thus in
order to obtain many more bright satellites than are observed in the
MW, it is clear that these satellites must have surface brightness
much dimmer than the known bright MW satellites.

\section{Comparison to Previous Results} 

There have been several recent analyses 
on the population of bright satellites around MW-analog galaxies along the lines
presented in this paper. It is instructive to compare the results presented
here to these previous analyses. 

~\citet{Guo:2011qc} used SDSS DR7 to construct the luminosity function
of satellites down to the magnitude scale of Fornax, correcting for
the incompleteness of SDSS.  These authors used best-fitting
photometric redshifts from DR7 to eliminate obvious
background galaxies. Our analysis differs from these authors in that
we utilize both DR8 imaging and a maximum likelihood method that 
incorporates full photometric redshift probability distributions. We also directly 
quantify the bias in abundance counts for faint satellites that is incurred 
when utilizing available photometric redshifts.
Via somewhat different methods for cutting background galaxies, 
~\citet{Lares2011} use DR7 data to obtain a mean number of satellites down to the 
magnitude of Sagittarius for projected radii $\gtrsim 100$ kpc. As we discuss
above, we have verified that our results are consistent with these authors 
over the radial range considered, and further that we do not incur a significant
bias by including galaxies within projected radii $< 100$ kpc. 
~\citet{Tollerud:2011wt} utilize the DR7
volume-limited spectroscopic sample and find that $\sim 40\%$ of
MW-analogs have satellites brighter than the LMC within 250 kpc. 
~\citet{James2011} use H$\alpha$ narrow band imaging to
search for start forming galaxies around 143 spiral galaxies like the
MW, and find that nearly two-thirds do not have satellites that
resemble the Magellanic Clouds. These latter two results are consistent 
with the spectroscopic results that we present for bright satellites. 

\section{Discussion and Conclusion} 
\label{sec:conclusion} 
We have used DR8 photometric redshift data to limit the mean number of
satellites around MW-analog galaxies down to ten magnitudes fainter
than the MW.  At least down to the scale of Sagittarius, the results
indicate that the MW is not a significant statistical outlier in its
number of bright, classical satellites.

Our 90\% c.l. upper bound of $\lesssim 13$ satellites brighter than
the Fornax dSph already places a strict bound on the efficiency of
galaxy formation at the dSph luminosity scale. This is particularly
true considering that there are anywhere from $\sim 25-75$ dark matter
subhalos in the Aquarius simulations~\citep{Springel2008} that have
present-day circular velocities greater than that of Fornax.  Surface
brightness incompleteness could increase this number, but likely not
enough to bring it into agreement with predictions for the number of
dense satellites in simulations.  However, it is very interesting to
note that the observational result we present is perfectly consistent
with abundance matching extrapolations for the satellite luminosity
function, which predict $\sim 1.2, 1.7$ satellites for magnitude
differences $\Delta m = 7,10$ ~\citep{Busha:2010gi}.  This does not
guarentee that such models will have the correct velocity function; in
fact it appears increasingly difficult to simultaneously match both
the luminosities and velocities of all of the satellites down to the
Fornax scale.

In the future it will be exceedingly important to increase the sample
of primary galaxies around which it is possible to measure satellites
as faint as Fornax.  Measuring the magnitude distribution at this
faint scale will go a long way towards determining if the mapping
between bright dSphs and dark matter subhalos is revealing the
presence of detailed baryonic physics not yet accounted for in
numerical simulations~\citep{Parry:2011iz} or about the properties of
dark matter~\citep{Lovell:2011rd}.

The principle uncertainty in accurately determining the satellite
distribution from the SDSS photometric sample is the fidelity of the
photometric redshifts.  We have shown that our method using the full
photometric redshift distribution is in excellent agreement with a
method that directly uses spectroscopic redshifts for bright
satellites where the training sample is representative.  However, this
method is biased to lower satellite numbers for dimmer magnitudes,
which is an indication that the spectroscopic training samples used to
construct the photometric redshift distributions are systematically
missing the lowest redshift galaxies.  Our measurements would be
substantially improved if these samples were unbiased, as this would
allow us to use the $p(z)$ method for a substantially larger number of
host galaxies.  It may be possible in the very near future to extend
the training sample to include data from
GAMA\footnote{http://www.gama-survey.org/}, extending to $r<19.8$,
which would allow us to extend our measurement using the $p(z)$ method
to the Fornax scale.  With some care, it may also be possible to
refine and improve the PRIMUS redshift finder in the range relevant to
our low-redshift primaries ($z<0.02$), thereby providing a more reliable
photometric redshift training sample.

Given forthcoming data sets it will also be possible to significantly
increase the number of primary galaxies; for example, the Dark Energy Survey
(DES)\footnote{http://www.darkenergysurvey.org/} will produce
a survey over 5000 sq. degrees and will observe galaxies to 24.3 in
the $\imath$-band.  If one considers satellites down to $r=24$, the
DES is expected to identify more than 4000 primaries with satellites
to Fornax magnitude differences, about 1600 primaries with
Leo I-like satellites, and nearly 100 primaries with satellites as
dim relative to their primary as Sculptor is to the Milky Way.  At the
Fornax scale, the statistics should be large enough to get a solid
measurement using background subtraction even without robust
photometric redshifts.  The analysis here indicates that the primary
challenge for the dimmest satellites around these primaries will be
determining their redshift distribution.

In the future it will also be important to obtain kinematics on
spectroscopically-confirmed satellites, as well as those that have a
high probability to be satellites from their photometric
redshifts. Comparing the velocity dispersion of these satellites to
the $\sim 10$ km/s velocity dispersions of the MW dSphs will allow for
a determination of both the luminosity function and the mass function
of satellites down to the scale of classical dSphs. This is only
currently possible with the sample of MW dSphs.

\section*{Acknowledgements}
We thank Carlos Cunha for extensive discussions about photometric
redshifts and comments on a draft, Michael Blanton for helpful
discussions about the spectroscopic sample and surface brightness
completeness, Erin Sheldon for help with edge effects and completeness
issues for the photometric catalog, Brian Gerke and Lulu Liu for
providing code from our previous work, John Moustakas and Alison Coil
for helpful discussions about the PRIMUS catalog at low redshift,
Marla Geha for many helpful discussions, and Michael Busha and Beth Willman 
for helpful comments on a draft.  LES and RHW were supported by the National
Science Foundation under grant NSF AST-090883.  Part of this work was
completed at the Aspen Center for Physics, with support from the
National Science Foundation under Grant No. 1066293.

Funding for the NASA-Sloan Atlas has been provided by the NASA
Astrophysics Data Analysis Program (08-ADP08-0072).  Funding for
SDSS-III has been provided by the Alfred P. Sloan Foundation, the
Participating Institutions, the National Science Foundation, and the
U.S. Department of Energy. The SDSS-III web site is
http://www.sdss3.org.  SDSS-III is managed by the Astrophysical
Research Consortium for the Participating Institutions of the SDSS-III
Collaboration including the University of Arizona, the Brazilian
Participation Group, Brookhaven National Laboratory, University of
Cambridge, University of Florida, the French Participation Group, the
German Participation Group, the Instituto de Astrofisica de Canarias,
the Michigan State/Notre Dame/JINA Participation Group, Johns Hopkins
University, Lawrence Berkeley National Laboratory, Max Planck
Institute for Astrophysics, New Mexico State University, New York
University, Ohio State University, Pennsylvania State University,
University of Portsmouth, Princeton University, the Spanish
Participation Group, University of Tokyo, University of Utah,
Vanderbilt University, University of Virginia, University of
Washington, and Yale University.  The Galaxy Evolution Explorer
(GALEX) is a NASA Small Explorer. The mission was developed in
cooperation with the Centre National d'Etudes Spatiales of France and
the Korean Ministry of Science and Technology.

%\bibstyle{apj} 
%\bibliography{bib}

\begin{thebibliography}{37}
%\expandafter\ifx\csname natexlab\endcsname\relax\def\natexlab#1{#1}\fi

\bibitem[{Aihara {et~al.}(2011)}]{Aihara:2011sj}
Aihara, H., {et~al.} 2011, Astrophys.J.Suppl., 193, 29

\bibitem[{Benson {et~al.}(2002)Benson, Lacey, Baugh, Cole, \&
  Frenk}]{Benson:2001au}
Benson, A.~J., Lacey, C., Baugh, C., Cole, S., \& Frenk, C. 2002,
  Mon.Not.Roy.Astron.Soc., 333, 156

\bibitem[{{Blanton} {et~al.}(2005){Blanton}, {Lupton}, {Schlegel}, {Strauss},
  {Brinkmann}, {Fukugita}, \& {Loveday}}]{Blanton2005}
{Blanton}, M.~R., {Lupton}, R.~H., {Schlegel}, D.~J., {Strauss}, M.~A.,
  {Brinkmann}, J., {Fukugita}, M., \& {Loveday}, J. 2005, \apj, 631, 208

\bibitem[{Boylan-Kolchin {et~al.}(2011)Boylan-Kolchin, Bullock, \&
  Kaplinghat}]{BoylanKolchin:2011de}
Boylan-Kolchin, M., Bullock, J.~S., \& Kaplinghat, M. 2011,
  Mon.Not.Roy.Astron.Soc., 415, L40

\bibitem[{Boylan-Kolchin {et~al.}(2010)Boylan-Kolchin, Springel, White, \&
  Jenkins}]{BoylanKolchin:2009an}
Boylan-Kolchin, M., Springel, V., White, S.~D., \& Jenkins, A. 2010,
  Mon.Not.Roy.Astron.Soc., 406, 896

\bibitem[{Bullock {et~al.}(2000)Bullock, Kravtsov, \&
  Weinberg}]{Bullock:2000wn}
Bullock, J.~S., Kravtsov, A.~V., \& Weinberg, D.~H. 2000, Astrophys.J., 539,
  517

\bibitem[{Busha {et~al.}(2011)Busha, Wechsler, Behroozi, Gerke, Klypin,
  {et~al.}}]{Busha:2010gi}
Busha, M.~T., Wechsler, R.~H., Behroozi, P.~S., Gerke, B.~F., Klypin, A.~A.,
  {et~al.} 2011, Astrophys. J., in press

\bibitem[{{Cannon} {et~al.}(2006){Cannon}, {Drinkwater}, {Edge}, {Eisenstein},
  {Nichol}, {Outram}, {Pimbblet}, {de Propris}, {Roseboom}, {Wake}, {Allen},
  {Bland-Hawthorn}, {Bridges}, {Carson}, {Chiu}, {Colless}, {Couch}, {Croom},
  {Driver}, {Fine}, {Hewett}, {Loveday}, {Ross}, {Sadler}, {Shanks}, {Sharp},
  {Smith}, {Stoughton}, {Weilbacher}, {Brunner}, {Meiksin}, \&
  {Schneider}}]{Cannon2006}
{Cannon}, R., {et~al.} 2006, \mnras, 372, 425

\bibitem[{Chen {et~al.}(2006)Chen, Kravtsov, Prada, Sheldon, Klypin,
  {et~al.}}]{Chen:2005sg}
Chen, J., Kravtsov, A.~V., Prada, F., Sheldon, E.~S., Klypin, A.~A., {et~al.}
  2006, Astrophys.J., 647, 86

\bibitem[{Coil {et~al.}(2011)Coil, Blanton, Burles, Cool, Eisenstein,
  {et~al.}}]{Coil:2010df}
Coil, A.~L., Blanton, M.~R., Burles, S.~M., Cool, R.~J., Eisenstein, D.~J.,
  {et~al.} 2011, Astrophys.J., 741, 8

\bibitem[{Cooper {et~al.}(2010)Cooper, Cole, Frenk, White, Helly,
  {et~al.}}]{Cooper:2009kx}
Cooper, A., Cole, S., Frenk, C., White, S., Helly, J., {et~al.} 2010,
  Mon.Not.Roy.Astron.Soc., 406, 744

\bibitem[{{di Cintio} {et~al.}(2011){di Cintio}, {Knebe}, {Libeskind}, {Yepes},
  {Gottl{\"o}ber}, \& {Hoffman}}]{DiCintio2011}
{di Cintio}, A., {Knebe}, A., {Libeskind}, N.~I., {Yepes}, G., {Gottl{\"o}ber},
  S., \& {Hoffman}, Y. 2011, \mnras, 417, L74

\bibitem[{Diemand {et~al.}(2008)Diemand, Kuhlen, Madau, Zemp, Moore,
  {et~al.}}]{Diemand:2008in}
Diemand, J., Kuhlen, M., Madau, P., Zemp, M., Moore, B., {et~al.} 2008, Nature,
  454, 735

\bibitem[{{Font} {et~al.}(2011){Font}, {Benson}, {Bower}, {Frenk}, {Cooper},
  {De Lucia}, {Helly}, {Helmi}, {Li}, {McCarthy}, {Navarro}, {Springel},
  {Starkenburg}, {Wang}, \& {White}}]{Font:2011ds}
{Font}, A.~S., {et~al.} 2011, \mnras, 417, 1260

\bibitem[{{Garilli} {et~al.}(2008){Garilli}, {Le F{\`e}vre}, {Guzzo},
  {Maccagni}, {Le Brun}, {de la Torre}, {Meneux}, {Tresse}, {Franzetti},
  {Zamorani}, {Zanichelli}, {Gregorini}, {Vergani}, {Bottini}, {Scaramella},
  {Scodeggio}, {Vettolani}, {Adami}, {Arnouts}, {Bardelli}, {Bolzonella},
  {Cappi}, {Charlot}, {Ciliegi}, {Contini}, {Foucaud}, {Gavignaud}, {Ilbert},
  {Iovino}, {Lamareille}, {McCracken}, {Marano}, {Marinoni}, {Mazure},
  {Merighi}, {Paltani}, {Pell{\`o}}, {Pollo}, {Pozzetti}, {Radovich}, {Zucca},
  {Blaizot}, {Bongiorno}, {Cucciati}, {Mellier}, {Moreau}, \&
  {Paioro}}]{Garilli2008}
{Garilli}, B., {et~al.} 2008, \aap, 486, 683

\bibitem[{{Guo} {et~al.}(2011){Guo}, {Cole}, {Eke}, \& {Frenk}}]{Guo:2011qc}
{Guo}, Q., {Cole}, S., {Eke}, V., \& {Frenk}, C. 2011, \mnras, 417, 370

\bibitem[{{James} \& {Ivory}(2011)}]{James2011}
{James}, P.~A., \& {Ivory}, C.~F. 2011, \mnras, 411, 495

\bibitem[{{Kleyna} {et~al.}(1997){Kleyna}, {Geller}, {Kenyon}, \&
  {Kurtz}}]{1997AJ....113..624K}
{Kleyna}, J.~T., {Geller}, M.~J., {Kenyon}, S.~J., \& {Kurtz}, M.~J. 1997, \aj,
  113, 624

\bibitem[{{Lares} {et~al.}(2011){Lares}, {Lambas}, \&
  {Dom{\'{\i}}nguez}}]{Lares2011}
{Lares}, M., {Lambas}, D.~G., \& {Dom{\'{\i}}nguez}, M.~J. 2011, \aj, 142, 13

\bibitem[{Lilly {et~al.}(2007)}]{Lilly:2006va}
Lilly, S., {et~al.} 2007, Astrophys.J.Suppl., 172, 70

\bibitem[{{Lilly} {et~al.}(1995){Lilly}, {Le Fevre}, {Crampton}, {Hammer}, \&
  {Tresse}}]{Lilly1995}
{Lilly}, S.~J., {Le Fevre}, O., {Crampton}, D., {Hammer}, F., \& {Tresse}, L.
  1995, \apj, 455, 50

\bibitem[{Liu {et~al.}(2011)Liu, Gerke, Wechsler, Behroozi, \&
  Busha}]{Liu:2010tn}
Liu, L., Gerke, B.~F., Wechsler, R.~H., Behroozi, P.~S., \& Busha, M.~T. 2011,
  Astrophys. J., 733, 62

\bibitem[{{Lovell} {et~al.}(2011){Lovell}, {Eke}, {Frenk}, {Gao}, {Jenkins},
  {Theuns}, {Wang}, {Boyarsky}, \& {Ruchayskiy}}]{Lovell:2011rd}
{Lovell}, M., {et~al.} 2011, arXiv:1104.2929

\bibitem[{Mandelbaum {et~al.}(2008)Mandelbaum, Seljak, Hirata, Bardelli,
  Bolzonella, {et~al.}}]{Mandelbaum:2007dp}
Mandelbaum, R., Seljak, U., Hirata, C., Bardelli, S., Bolzonella, M., {et~al.}
  2008, Mon.Not.Roy.Astron.Soc., 386, 781

\bibitem[{{Parry} {et~al.}(2011){Parry}, {Eke}, {Frenk}, \&
  {Okamoto}}]{Parry:2011iz}
{Parry}, O.~H., {Eke}, V.~R., {Frenk}, C.~S., \& {Okamoto}, T. 2011,
  arXiv:1105.3474

\bibitem[{Sheldon {et~al.}(2011)Sheldon, Cunha, Mandelbaum, Brinkmann, \&
  Weaver}]{Sheldon:2011fm}
Sheldon, E.~S., Cunha, C., Mandelbaum, R., Brinkmann, J., \& Weaver, B.~A.
  2011, arXiv:1109.5192

\bibitem[{{Somerville}(2002)}]{Somerville:2001km}
{Somerville}, R.~S. 2002, \apjl, 572, L23

\bibitem[{{Springel} {et~al.}(2008){Springel}, {Wang}, {Vogelsberger},
  {Ludlow}, {Jenkins}, {Helmi}, {Navarro}, {Frenk}, \& {White}}]{Springel2008}
{Springel}, V., {et~al.} 2008, \mnras, 391, 1685

\bibitem[{{Strigari} {et~al.}(2010){Strigari}, {Frenk}, \&
  {White}}]{Strigari:2010un}
{Strigari}, L.~E., {Frenk}, C.~S., \& {White}, S.~D.~M. 2010, \mnras, 408, 2364

\bibitem[{Tollerud {et~al.}(2011)Tollerud, Boylan-Kolchin, Barton, Bullock, \&
  Trinh}]{Tollerud:2011wt}
Tollerud, E.~J., Boylan-Kolchin, M., Barton, E.~J., Bullock, J.~S., \& Trinh,
  C.~Q. 2011, Astrophys.J., 738, 102

\bibitem[{{van den Bergh}(2000)}]{vandenBergh2000}
{van den Bergh}, S. 2000, \pasp, 112, 529

\bibitem[{{Wadepuhl} \& {Springel}(2011)}]{Wadepuhl2011}
{Wadepuhl}, M., \& {Springel}, V. 2011, \mnras, 410, 1975

\bibitem[{{Walsh} {et~al.}(2009){Walsh}, {Willman}, \& {Jerjen}}]{Walsh2009}
{Walsh}, S.~M., {Willman}, B., \& {Jerjen}, H. 2009, \aj, 137, 450

\bibitem[{Weiner {et~al.}(2005)Weiner, Phillips, Faber, Willmer, Vogt,
  {et~al.}}]{Weiner:2004qm}
Weiner, B.~J., Phillips, A.~C., Faber, S., Willmer, C.~N., Vogt, N.~P.,
  {et~al.} 2005, Astrophys.J., 620, 595

\bibitem[{{Willman}(2010)}]{2010AdAst2010E..21W}
{Willman}, B. 2010, Advances in Astronomy, 2010

\bibitem[{Wirth {et~al.}(2004)Wirth, Willmer, Amico, Chaffee, Goodrich,
  {et~al.}}]{Wirth:2004tm}
Wirth, G.~D., Willmer, C.~N., Amico, P., Chaffee, F.~H., Goodrich, R.~W.,
  {et~al.} 2004, Astron.J., 127, 3121

\bibitem[{Yee {et~al.}(2000)Yee, Morris, Lin, Carlberg, Hall,
  {et~al.}}]{Yee:2000qj}
Yee, H., Morris, S., Lin, H., Carlberg, R., Hall, P., {et~al.} 2000,
  Astrophys.J.Suppl., 129, 475

\end{thebibliography}

\end{document}